\documentclass[a4paper,twocolumn,english,aps,prb,floatfix,showpacs,preprintnumbers]{revtex4}
\usepackage[latin9]{inputenc}
\usepackage{color}
\usepackage{array}
\usepackage{multirow}
\usepackage{amsmath}
\usepackage{amssymb}
\usepackage{graphicx}
\usepackage{esint}

\makeatletter

\providecommand{\tabularnewline}{\\}

\@ifundefined{textcolor}{}
{%
 \definecolor{BLACK}{gray}{0}
 \definecolor{WHITE}{gray}{1}
 \definecolor{RED}{rgb}{1,0,0}
 \definecolor{GREEN}{rgb}{0,1,0}
 \definecolor{BLUE}{rgb}{0,0,1}
 \definecolor{CYAN}{cmyk}{1,0,0,0}
 \definecolor{MAGENTA}{cmyk}{0,1,0,0}
 \definecolor{YELLOW}{cmyk}{0,0,1,0}
}

\usepackage{babel}

\usepackage{babel}

\usepackage{babel}

\usepackage{babel}

\usepackage{babel}

\usepackage{babel}

\makeatother

\usepackage{babel}
\begin{document}

\title{On adatomic-configuration-mediated correlation between electrotransport
and electrochemical properties of graphene}

\author{T. M. Radchenko,$^{1}$ V. A. Tatarenko,$^{1}$ I. Yu. Sagalianov,$^{2}$
Yu. I. Prylutskyy,$^{2}$ P. Szroeder,$^{3}$ and S. Biniak$^{4}$}

\affiliation{$\mathit{\mathrm{\textrm{\ensuremath{^{1}}}}}$Deptartment of Solid
State Theory, G. V. Kurdyumov Institute for Metal Physics of NASU,
36 Acad. Vernadsky Blvd., UA-03680 Kyiv, Ukraine}

\affiliation{$^{2}$Taras Shevchenko National University of Kyiv, 64 Volodymyrska
Str., UA-03022 Kyiv, Ukraine}

\affiliation{$^{3}$Faculty of Physics, Astronomy and Informatics, Institute of
Physics, Nicolaus Copernicus University, Grudziadzka 5/7, 87-100 Toru\'{n},
Poland}

\affiliation{$^{4}$Faculty of Chemistry, Nicolaus Copernicus University, Gagarina
7, 87-100 Toru\'{n}, Poland}

\date{\today }
\begin{abstract}
The electron-transport properties of adatom--graphene system are investigated
for different spatial configurations of adsorbed atoms: when they
are randomly-, correlatively-, or orderly-distributed over different
types of high symmetry sites with various adsorption heights. Potassium
adatoms in monolayer graphene are modeled by the scattering potential
adapted from the independent self-consistent \textit{ab initio} calculations.
The results are obtained numerically using the quantum-mechanical
Kubo--Greenwood formalism. A band gap may be opened only if ordered
adatoms act as substitutional atoms, while there is no band gap opening
for adatoms acting as interstitial atoms. The type of adsorption sites
strongly affect the conductivity for random and correlated adatoms,
but practically does not change the conductivity when they form ordered
superstructures with equal periods. Depending on electron density
and type of adsorption sites, the conductivity for correlated and
ordered adatoms is found to be enhanced in dozens of times as compared
to the cases of their random positions. These the correlation and
ordering effects manifest weaker or stronger depending on whether
adatoms act as substitutional or interstitial atoms. The conductivity
approximately linearly scales with adsorption height of random or
correlated adatoms, but remains practically unchanged with adequate
varying of elevation of ordered adatoms. Correlations between electron
transport properties and heterogeneous electron transfer kinetics
through potassium-doped graphene and electrolyte interface are investigated
as well. The ferri-/ferrocyanide redox couple is used as an electrochemical
benchmark system. Potassium adsorption of graphene electrode results
to only slight suppress of the heterogeneous standard rate constant.
Band gap, opening for ordered and strongly short-range scatterers,
has a strong impact on the dependence of the electrode reaction rate
as a function of electrode potential. 
\end{abstract}


\maketitle

\section{Introduction}

Adsorbed atoms and molecules are probably the most important examples
of point defects in the physics of graphene.\cite{Katsnelson2012}
In addition to remarkable intrinsic electronic and mechanical properties
of pure graphene, its structure and properties can also be modified
and controlled by adsorption and doping of atoms and molecules. That
is why last few years studies of atom adsorption of both metallic
\cite{Lugo-Solis2007,Uchoa2008,Gan2008,Chan2008,Mao2008,Sevincli2008,Zanella2008,Chen2008,Ishii2008,Malola2009,Wu2009,Wehling2009,Wang2009,Uthaisar2009,Krasheninnikov2009,Khomyakov2009,Johll2009,CastroNeto2009,Martinez2009,Suarez-Martinez2009,Vo-Van2010,Kubota2010,Klintenberg2010,Jin2010,Zolyomi2010,Rodriguez-Manzo2010,Santos2010,Cao2010,Valencia2010,Qiao2010,Hupalo2011,Liu2010,Liu2011,Liu2012,Nakada2011,Sachs2011,McCreary2011,Hardcastle2013,Garcia2014,Virgus2014,Sessi2014,Naji2014,Jia2015}
and nonmetallic \cite{Ishii2008,Wehling2009,Yuan2010,Klintenberg2010,Ataca2011,Nakada2011,Sachs2011,Wu2008,Zhou2009,Gao2010,Ivanovskaya2010,Lin2015}
adsorbates on graphene attract a considerable attention. Overwhelming
majority of theoretical and computational studies of adatom--graphene
systems deal with first-principles density-functional calculations,
which require high computational capabilities, therefore the size
of graphene computational domains in these calculations are mostly
limited to periodic supercells and localized fragments containing
a relatively small number of atoms (sites). Nevertheless, the first-principle
study is suitable and fruitful, and therefore prevalent now, for calculation
of energetic, structural, and magnetic parameters: adsorption (binding)
energy and height of adatoms, diffusion (migration) barrier energy,
in-plane and vertical graphene-lattice distortion amplitude, charge
transfer, electric-dipole moment, magnetic moments of an isolated
atom and total graphene--adatom system, \textit{etc}. \cite{Lugo-Solis2007,Uchoa2008,Chan2008,Mao2008,Sevincli2008,Zanella2008,Ishii2008,Wehling2009,Khomyakov2009,Martinez2009,Wang2009,Johll2009,Wu2009,Krasheninnikov2009,Malola2009,Suarez-Martinez2009,Santos2010,Cao2010,Valencia2010,Liu2010,Liu2011,Liu2012,Klintenberg2010,Jin2010,Zolyomi2010,Qiao2010,Hupalo2011,Ataca2011,Nakada2011,Sachs2011,Hardcastle2013,Naji2014,Wu2008,Zhou2009,Gao2010,Ivanovskaya2010,Lin2015} 

\begin{figure*}
\includegraphics[width=0.95\textwidth]{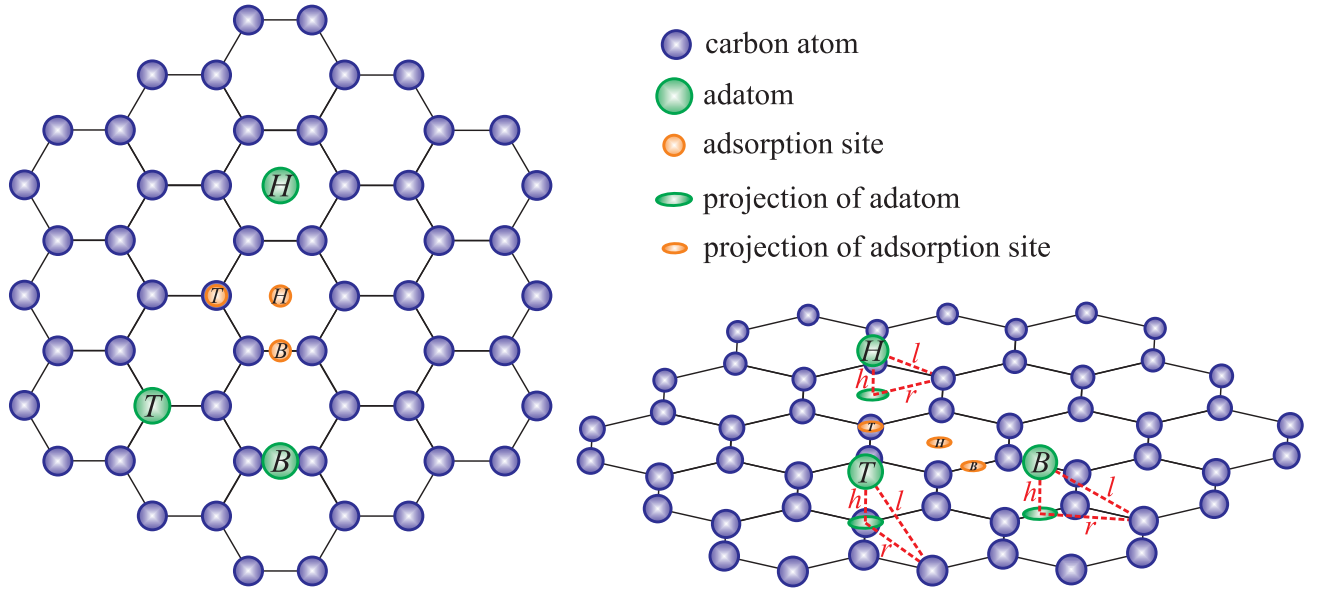}

\caption{(Color online) Typical configurations of adatom--graphene system:
top (left) and perspective (right) views of graphene lattice with
hollow center (\textit{H}), bridge center (\textit{B}), and (a)top
(\textit{T}) adsorption sites.}

\label{Fig_Adatoms} 
\end{figure*}

Because of the hexagonal symmetry of the graphene lattice, possible
adsorption sites for a single atom can be reduced into three types
with high-symmetry positions: so-called hollow center (\textit{H}-type),
bridge center (\textit{B}-type), and (a)top (\textit{T}-type) adsorption
sites as illustrated in Fig.~\ref{Fig_Adatoms}. The most favorable
(stable) adsorption site is determined by placing the adatom onto
these three adsorption sites, and each time by optimizing structures
to obtain minimum energy and atomic forces; as a result, the highest
binding (adsorption) energy of adatom corresponds to its the most
favorable site. Analysis of the density-functional-theory-based studies,
\cite{Lugo-Solis2007,Uchoa2008,Chan2008,Mao2008,Sevincli2008,Zanella2008,Ishii2008,Wehling2009,Khomyakov2009,Martinez2009,Wang2009,Johll2009,Wu2009,Krasheninnikov2009,Malola2009,Suarez-Martinez2009,Santos2010,Cao2010,Valencia2010,Liu2010,Liu2011,Liu2012,Klintenberg2010,Jin2010,Zolyomi2010,Qiao2010,Hupalo2011,Ataca2011,Nakada2011,Sachs2011,Hardcastle2013,Naji2014}
covering almost all the periodic table, yields: (i) for metals, the
most stable adsorption sites are the \textit{H-}sites, followed by
the \textit{B-}sites, and then the \textit{T}-sites, although the
energy differences between the \textit{H} and \textit{B} or \textit{T}
sites are very small for the alkali and group-III metals, particularly
for potassium (see Table~\ref{Table_potassium}), which we regard
as an example of adsorbate in the present study; (ii) for both metals
and nonmetals, adsorption heights for more favorable sites are lower
as compared with heights of the lesser favorable adsorption sites. 

Data of Table~\ref{Table_potassium} for K adsorption on graphene
read that values of adsorption energy reported in the literature disagree
by as much as almost 100\%, while adsorption heights differ by up
to 5\%.\cite{Note1} Similar inconsistencies of the literature data
occur also for other periodic-table elements. For example, Cu and
Sn prefer \textit{T}-site bonding (at the heights of 2.12~Å and 2.82~Å,
respectively) according to Refs. \onlinecite{Wu2009} and \onlinecite{Chan2008},
while \textit{B}-site bonding (2.03~Å and 2.42~Å) in accordance
with Refs. \onlinecite{Cao2010} and \onlinecite{Nakada2011}.
On the one hand, such discrepancies in determination of the energy
stability of adsorption sites has resulted in a controversy and questions
concerning the accuracy of theoretical models (calculations) used
in those studies. On the other hand, this motivates us to study how
the positioning of adatoms on each of \textit{H}, \textit{B}, and
\textit{T} site types affects the transport properties of graphene
in comparison with the cases of their location on two other types
of the sites. 

\begin{table}
\caption{Literature data on calculated adsorption energies and heights for
K adatoms occupying hollow (\textit{H}), bridge (\textit{B}), and
top (\textit{T}) adsorption sites in graphene.}

\begin{tabular}{llll}
\hline 
\multirow{2}{*}{Calculated parameter} & \multicolumn{3}{c}{Adsorption site }\tabularnewline
\cline{2-4} 
 & \textit{H}-type & \textit{B}-type & \textit{T}-type\tabularnewline
\hline 
\multirow{4}{*}{Adsorption energy {[}eV{]}} & 0.785$^{\mathrm{a}}$ & 0.726$^{\mathrm{a}}$ & 0.720$^{\mathrm{a}}$\tabularnewline
 & 0.802$^{\mathrm{b}}$ & 0.739$^{\mathrm{b}}$ & 0.733$^{\mathrm{b}}$\tabularnewline
 & 1.461$^{\mathrm{c}}$ & 1.403$^{\mathrm{c}}$ & 1.405$^{\mathrm{c}}$\tabularnewline
 & 0.810$^{\mathrm{d}}$ &  & \tabularnewline
\hline 
\multirow{4}{*}{Adsorption height {[}Å{]}} & 2.62$^{\mathrm{a}}$ &  & \tabularnewline
 & 2.60$^{\mathrm{b}}$ & 2.67$^{\mathrm{b}}$ & 2.67$^{\mathrm{b}}$\tabularnewline
 & 2.52$^{\mathrm{c}}$ & 2.59$^{\mathrm{c}}$ & 2.55$^{\mathrm{c}}$\tabularnewline
 & 2.58$^{\mathrm{d}}$ &  & \tabularnewline
\hline 
\end{tabular}

\begin{raggedright}
\label{Table_potassium} 
\par\end{raggedright}

\raggedright{}$^{\mathrm{a\mathrm{-}d}}$References~\onlinecite{Liu2011,
Chan2008, Qiao2010, Nakada2011} (respectively).
\end{table}

Distributions of adatoms over the \textit{H}, \textit{B}, or \textit{T}
graphene-lattice adsorption sites are not always random, as it is
usually in three-dimensional metals and alloys, where adatoms are
introduced by alloying, which is generically a random process.\cite{CastroNeto2009}
Diluted atoms may have a tendency towards the spatial correlation\cite{YanFuhrer2011}
or even ordering.\cite{Cheianov2010,Cheianov2009PRB,Cheianov2009SSC,Howard2011,Song2012,Lin2015}
Moreover, since graphene is an open surface, (ad)atoms can be positioned
onto it with the use of scanning tunneling\cite{Eigler1990} or transmission
electron\cite{Meyer2008} microscopes allowing to design (ad)atomic
configurations as well as ordered (super)structures with atomic precision.
Recently, several ordered configurations of hydrogen adatoms on graphene
have been already directly observed by scanning tunneling microscopy
in Ref.~\onlinecite{Lin2015}. 

Though many properties of atom adsorption onto graphene have been
extensively studied in many works, there is still no one paper on
how such a variety of the spatial arrangements of adatoms (\textit{viz}.,
their random, correlated, and ordered distributions in the \textit{H},
\textit{B}, and \textit{T} types of bonding with varying adsorption
heights) influences (if any) on electron transport in graphene. Such
a problem formulation arises in context of the possibility to consider
(ad)atomic spatial configurations as an additional tool for modification
and controlling graphene's transport properties. 

Another part of our paper deals with attempt to detect adatom-mediated
correlation between electron transport and electrochemical properties
of graphene. Understanding of its electrochemical properties, especially
the electron transfer kinetics of a redox reaction between graphene
surface (electrode) and redox couple in electrolyte, is essential\cite{Velicky2014}
for its potential in energy conversion and storage to be realized,\cite{Zhao2009,Bonacorso2015}
as well as opens up interesting opportunities for using graphene as
an electrode material for field effect transistors\cite{Li2009,Yan2011}
and electrochemical senors.\cite{Ratinac2010,Lia2012} To examine
the heterogeneous electron transfer kinetics at highly oriented pyrolytic
graphite (HOPG) and glassy carbon (GC) electrode, several electroactive
species were used.\cite{McCreery2008} Results show that electron
transfer is slower at the basal plane of HOPG than at the edge plane.
The kinetics of the electron transfer is enhanced after electrode
pretreatment. However, in epitaxial graphene, only a part of the surface
is electroactive, even after electrochemical pretreatment.\cite{Szroeder2014}
Experimental results confirmed the belief that point and edge defects
as well as oxygenated functional groups can mediate electron transfer.\cite{Cline1994}
Contrary to the traditional view, high-resolution electrochemical
imaging experiments have revealed that electron transfer occurs at
both the basal planes of graphite as at the edge sites.\cite{Lai2012}
To examine these discrepancies, we calculate the electron transfer
kinetics at graphene with randomly-, correlatively-, and orderly-adsorbed
atoms described by scattering potential \textcolor{black}{manifesting
both short- and long-range features,} and also use strongly short-range
scattering potential. Results show that electron transfer still occurs
for adsorbed graphene.

The rest of the paper is organized as follows. Section~II consists
of two subsections containing models for electron transport and transfer.
In the first subsection, we formulate the Kubo--Greenwood-formalism-based
numerical model for electron transport in graphene, which is appropriate
for realistic graphene sheets with millions of atoms. The size of
our computational domain is up to 10 millions of atoms that corresponds
to $\approx500\times500$~nm$^{2}$. The second subsection encloses
the basic model we use to calculate the rate constant of electron
transfer between solid (graphene) electrode and redox couple in electrolyte
using the Gerischer--Marcus approach. Section~III presents and discusses
the obtained results. Finally, the conclusions of our work are given
in Sec.~IV.

\section{Models}

\subsection{Electron transport}

To investigate the charge transport in adatom--graphene system, an
exact numerical technique within the Kubo--Greenwood formalism, \cite{Madelung1996,Roche1997,Roche2012,Markussen2006,Triozon2002,Triozon2004,Lherbier2008_1,Lherbier2008_2,Lherbier2011,Leconte2011,Lherbier2012,Trambly2011,Ishii2010,Tuan2013,Radchenko_PRB2012,Radchenko_PRB2013,Radchenko_SSC2014,Radchenko_PLA2014,Torres2014,Radchenko_NOVA2014}
which captures all (ballistic, diffusive, and localization) transport
regimes, is employed. Within the framework of this approach, the energy
($E$) and time ($t$) dependent diffusivity $D(E,t)$\cite{Note2}
is governed by the wave-packet propagation: \cite{Roche1997,Roche2012,Markussen2006,Triozon2002,Triozon2004,Lherbier2008_1,Lherbier2008_2,Lherbier2011,Leconte2011,Lherbier2012,Trambly2011,Ishii2010,Tuan2013,Radchenko_PRB2012,Radchenko_PRB2013,Radchenko_SSC2014,Radchenko_PLA2014,Torres2014,Radchenko_NOVA2014}
$D(E,t)=\bigl\langle\Delta\hat{X}^{2}(E,t)\bigr\rangle/t$,\cite{Note3}
where the mean quadratic spreading of the wave packet along the direction
$x$ reads as \cite{Roche1997,Roche2012,Markussen2006,Triozon2002,Triozon2004,Lherbier2008_1,Lherbier2008_2,Lherbier2011,Leconte2011,Lherbier2012,Trambly2011,Ishii2010,Tuan2013,Radchenko_PRB2012,Radchenko_PRB2013,Radchenko_SSC2014,Radchenko_PLA2014,Torres2014,Radchenko_NOVA2014} 

\begin{equation}
\bigl\langle\Delta\hat{X}^{2}(E,t)\bigr\rangle=\frac{\text{Tr}[\hat{(X}(t)-\hat{X}(0))^{2}\delta(E-\hat{H})]}{\text{Tr}[\delta(E-\hat{H})]}\label{Eq_Spreading}
\end{equation}
with $\hat{X}(t)=\hat{U}^{\dagger}(t)\hat{X}\hat{U}(t)$---the position
operator in the Heisenberg representation, $\hat{U}(t)=e^{-i\hat{H}t/\hbar}$---the
time-evolution operator, and a standard $p$-orbital nearest-neighbor
tight-binding Hamiltonian $\hat{H}$ is \cite{PeresReview2010,DasSarmaReview2011}

\begin{equation}
\hat{H}=-u{\textstyle \sum_{i,i^{\prime}}}c_{i}^{\dagger}c_{i^{\prime}}+{\textstyle \sum_{i}}V_{i}c_{i}^{\dagger}c_{i},\label{Eq_Hamiltonian}
\end{equation}
where $c_{i}^{\dagger}$ ($c_{i}$) is a standard creation (annihilation)
operator acting on a quasiparticle at the site $i$. The summation
over $i$ runs the entire honeycomb lattice, while $i^{\prime}$ is
restricted to the sites next to $i$; $u=2.7$ eV is the hopping integral
for the neighboring C atoms occupying $i$ and $i^{\prime}$ sites
at a distance $a=0.142$ nm between them; and $V_{i}$ is the on-site
potential defining scattering strength on a given graphene-lattice
site $i$ due to the presence of impurity adatoms. The impurity scattering
potential plays a crucial role in the transport model we use at hand.

\begin{figure*}
\includegraphics[width=0.8\textwidth]{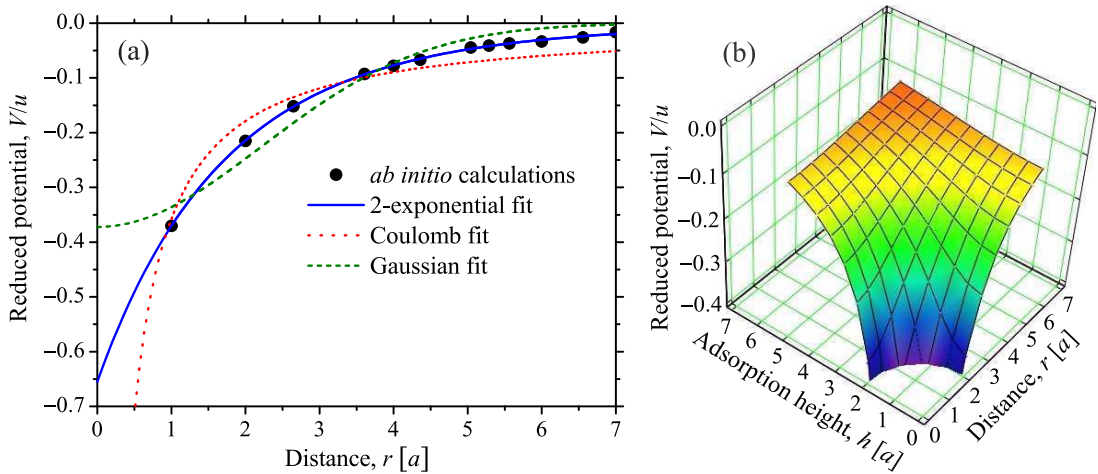}

\caption{(Color online) Scattering potential for potassium adatoms with (a)
fixed adsorption height $h=2.4$ Å\textcolor{black}{{} and (b) varying
$h$. Here, }\textit{\textcolor{black}{ab initio}}\textcolor{black}{{}
calculations }($\bullet$) \cite{Adessi2006} are fitted\cite{Fitting}
by different functions, \textit{viz}\textcolor{black}{. Gaussian ($V=Ue^{-\mathbf{\mathrm{\mathit{r}}}^{2}/2\xi^{2}}$
with fitting parameters $U=-0.37u$ and $\xi=2.21a$ defining a potential
height and an effective potential radius, respectively), Coulomb ($V=Q/r$
with $Q=-0.36ua$), and two-exponential ($V=U_{1}e^{-r/\xi_{1}}+U_{2}e^{-r/\xi_{2}}$
with $U_{1}=-0.45u$, $\xi_{1}=1.47a$, $U_{2}=-0.20u$, $\xi_{2}=2.73a$);}
$r$ is a distance from the projection of adatom to the lattice site
as shown in Fig.~\ref{Fig_Adatoms}.}

\label{Fig_potential} 
\end{figure*}

For adatoms located on \textit{H}-type sites (see Fig.~\ref{Fig_Adatoms}),
the impurity scattering potential in the Hamiltonian matrix is introduced
as on-site energies $V_{i}$ varying with distance $r$ to the center
of a hexagon on which the adatom projects according to the potential
profile $V=V(r)<0$ in Fig.~\ref{Fig_potential}(a) adapted from
the self-consistent \textit{ab initio} calculations\cite{Adessi2006}
for K adatoms on the height $h\cong2.4$ Å over the graphene surface.
As the fitting\cite{Fitting} shows, this potential is far from the
Coulomb- or Gaussian-like shapes commonly used in the literature for
charged impurities in graphene, \textcolor{black}{while two-exponential
fitting exactly reproduces the potential. Such a scattering potential
presents both short- and long-range features,\cite{Lherbier2008_2}
although its short-range characteristics become rather stronger for
adatoms that are nonrandom (correlated and ordered) in their spatial
positions. }Transforming scattering potential $V=V(r)$ into its dependence
on distance from the lattice site directly to adatom, $V=V(l)$, where
$l=\sqrt{r^{2}+h^{2}}$ as demonstrably from Fig.~\ref{Fig_Adatoms},
one can obtain its dependence on both $r$ and $h$, $V=V(r,h)$,
which is presented in Fig.~\ref{Fig_potential}(b). As follows from
Fig.~\ref{Fig_potential}(b), if $r=a$ and $h=2.4$ Å $=1.69a$,
$V=-0.37u$, which agrees with Fig.~\ref{Fig_potential}(a).

For adatoms positioned on \textit{B}- and \textit{T}-type sites (Fig.~\ref{Fig_Adatoms}),
we use the same scattering potential $V=V(r)$ as in Fig.~\ref{Fig_potential}(a)
with difference that $r$ denotes distance from the lattice site to
the middle of a C--C bond and to a C atom, respectively. Strictly
speaking, $V=V(r)$ for adatoms on \textit{H}-, \textit{B}-, and \textit{T}-type
sites (Fig.~\ref{Fig_Adatoms}) should be different, however just
approach of the same scattering potential for these three types of
adatom locations allows us to reveal manifestation of configurational
effects in the transport properties of graphene we are interested
in the present study. 

In case of correlation, adatoms are no longer considered to be randomly
located. To describe their spatial correlation, we adopt a model\cite{Li2011,Li2012}
using the pair distribution function $p(\mathbf{R}_{i}-\mathrm{\mathbf{R}}_{j})\equiv p(r)$:

\begin{equation}
p(r)=\left\{ \begin{array}{l}
0,\; r<r_{0}\\
1,\; r\geq r_{0}
\end{array}\right.\label{p_corr}
\end{equation}
where $r=|\mathbf{R}_{i}-\mathbf{R}_{j}|$ is a distance between the
two adatoms, and a correlation length $r_{0}$ defines minimal distance
that can separate any two of them. Note that for the randomly distributed
adatoms, $r_{0}=0$. Although the correlation length $r_{0}$ is found
to be insensitive to impurity (potassium) density,\cite{YanFuhrer2011}
the maximal correlation length $r_{0_{\mathrm{max}}}$ depends on
both relative adatom concentration and positions of adsorption sites
as given in Table~\ref{Table_correlation}. In our calculations for
$n_{\mathrm{K}}=3.125\%$ of correlated (potassium) adatoms, we chose
$r_{0}=r_{0_{\mathrm{max}}}^{H,B}=7a$ for hollow and bridge sites,
and $r_{0_{\mathrm{max}}}^{T}=5a$ for top sites. 

\begin{figure*}
\includegraphics[width=1\textwidth]{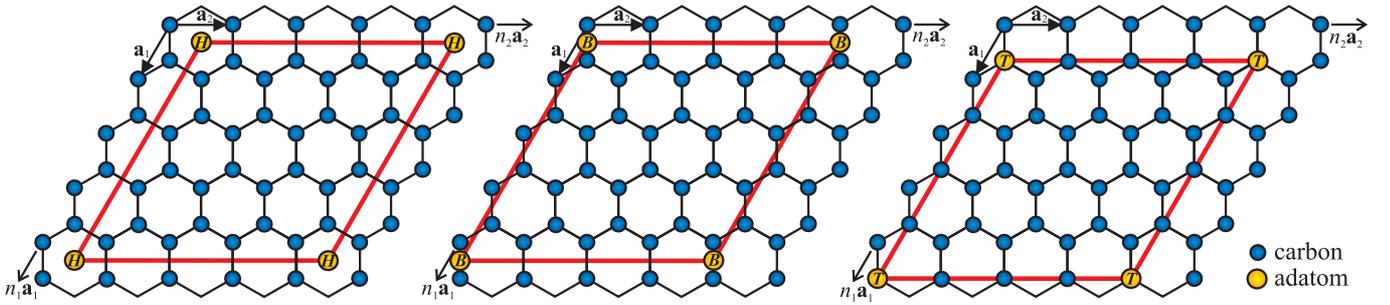}

\caption{(Color online) Top view of graphene lattice with ordered adatoms resided
on hollow (left), bridge (centre), and top (right) sites.}

\label{Fig_Ordering} 
\end{figure*}

In case of adatom ordering, we consider superlattice structures in
Fig.~\ref{Fig_Ordering}, where the relative content of ordered (potassium)
adatoms is the same as for random and correlation cases, $n_{\mathrm{K}}=1/32=3.125\%$.
This structures form interstitial {[}Fig.~\ref{Fig_Ordering}(left)
and Fig.~\ref{Fig_Ordering}(center){]} or substitutional {[}Fig.~\ref{Fig_Ordering}(right){]}
superstructures, where distribution of adatoms over the honeycomb-lattice
interstices or sites, respectively, can be described by the single-site
occupation-probability functions derived via the static concentration
wave method. \cite{Khachaturyan2008,Radchenko-SSP2009,Radchenko-SSS2009,Radchenko-PhysE2010,Radchenko-SSP2008}
In the computer implementation, $n_{\mathrm{K}}=3.125\%$ of potassium
adatoms occupy sites within the same sublattice and can be described
via a single-site function:

\begin{equation}
P(\mathbf{R})=\left\{ \begin{array}{l}
1,\; n_{1}+n_{2}=4\mathbb{Z}\\
0,\;\mathrm{otherwise}
\end{array}\right.\label{P_ord}
\end{equation}
where $n_{1}$, $n_{2}$, and $\mathbb{Z}$ belong to the set of integers,
$n_{1}$ and $n_{2}$ denote coordinates of sites in an oblique coordinate
system formed by the basis translation vectors $\mathbf{a}_{1}$ and
$\mathbf{a}_{2}$ shown in Fig.~\ref{Fig_Ordering}, and $\mathbf{R}$
denotes origin position of the unit cell where the considered interstice
{[}Fig.~\ref{Fig_Ordering}(left) and Fig.~\ref{Fig_Ordering}(center){]}
or site {[}Fig.~\ref{Fig_Ordering}(right){]} resides.

\begin{table}
\caption{Relation between the relative concentration of impurity adatoms ($n_{i}$)
occupying \textit{H}-, \textit{B}-, or \textit{T}-type sites (see
Fig.~\ref{Fig_Adatoms}) and the largest correlation distance ($r_{0_{\max}}$)
expressed in units of the lattice parameter $a=0.142$ nm.}

\begin{tabular}{llllllll}
\hline 
Site  & $n_{i}$  & 0.5\%  & 1\%  & 2\%  & 3\%  & 4\%  & 5\% \tabularnewline
\hline 
\textit{H},\textit{ B}  & $r_{0_{\max}}^{H,B}$ {[}$a${]} & $18$  & $13$  & $9$  & $7$  & $6$  & $5$\tabularnewline
\hline 
\textit{T}  & $r_{0_{\max}}^{T}$ {[}$a${]}  & $13$  & $9$  & $6$  & $5$  & $4$  & $3$\tabularnewline
\hline 
\end{tabular}

\label{Table_correlation} 
\end{table}

The dc conductivity $\sigma$ can be extracted from the diffusivity
$D(E,t)$, when it saturates reaching the maximum value, $\lim_{t\rightarrow\infty}D(E,t)=D_{\max}(E)$,
and the diffusive transport regime occurs. Then the semiclassical
conductivity at a zero temperature is defined as \cite{Leconte2011,Lherbier2012}

\begin{equation}
\sigma=e^{2}\tilde{\rho}(E)D_{\max}(E),\label{Eq_sigmaMax}
\end{equation}
where $-e<0$ denotes the electron charge and $\tilde{\rho}(E)=\rho/\Omega=\textrm{Tr}[\delta(E-\hat{H})]/\Omega$
is the density of sates (DOS) per unit area $\Omega$ (and per spin).
The DOS is also used to calculate the electron density as $n_{e}(E)=\int_{-\infty}^{E}\tilde{\rho}(E)dE-n_{\text{ions}},$
where $n_{\text{ions}}=3.9\cdot10^{15}$ cm$^{-2}$ is the density
of the positive ions in the graphene lattice compensating the negative
charge of the $p$-electrons (at the neutrality (Dirac) point of pristine
graphene, $n_{e}(E)=0$). Combining the calculated $n_{e}(E)$ with
$\sigma(E)$, we compute the density dependence of the conductivity
$\sigma=\sigma(n_{e})$.

Note that we do not go into details of numerical calculations of DOS,
$D(E,t)$, and $\sigma$ since details of the computational methods,
we utilize here (Chebyshev method for solution of the time-dependent
Schrödinger equation, calculation of the first diagonal element of
the Green's function using continued fraction technique and tridiagonalization
procedure of the Hamiltonian matrix, averaging over the realizations
of impurity adatoms, sizes of initial wave packet and computational
domain, boundary conditions, \textit{etc}.) are given by Radchenko
\textit{et al}. \cite{Radchenko_PRB2012}

\subsection{Electron transfer}

To calculate the heterogeneous rate constant of electron transfer
from the reduced form of redox couple to graphene electrode, we used
Gerischer--Marcus model.\cite{Gerischer1960_p223,Gerischer1960_p325,Gerischer1961,Gerischer1970,Memming2001,Szroeder2011,Szroeder2013}
In this model it is assumed that electron transfer between solid electrode
and redox couple in electrolyte is much faster than reorientation
of the solvent molecules (diabatic representation). As a result, the
rate constant of the electrode reaction depends only on the electron
DOS in the solid and the distribution energy levels of the reduced
(oxidized) form, $W_{\mathrm{Red\,(Ox)}}$ in the solution. If the
vacuum energy as a reference energy level is chosen, the electrochemical
potential of electrons occupying energy levels of ions, $\bar{\mu}_{\mathrm{e,redox}}$,
is equivalent to the Fermi level of the redox couple in the solution,
$E_{\mathrm{\mathit{F},redox}}$.\cite{Gerischer1983} As oxidized
and reduced form interact with surrounding polar solvent in a different
way, energy levels of oxidized and reduced form are shifted each other
by $2\lambda$, where $\lambda$ is the reorganization energy.

In our calculations, we used the Gaussian distribution of the electronic
states of the reduced form given by \cite{Memming2001} 
\begin{equation}
P(E)=\frac{1}{\sqrt{4k_{B}T\lambda}}\exp\!\left[-\frac{\left(E-E_{\mathrm{\mathit{F},redox}}-\lambda\right)^{2}}{4k_{B}T\lambda}\right],\label{Eq_Gauss_distrib}
\end{equation}
where $k_{B}$ is the Boltzmann constant, $T$ is the absolute temperature.
The Fe(CN)$_{3}^{3-/4-}$ redox couple has been chosen as a benchmark
system. For ferri-/ferrocyanide redox couple, the $\lambda$ value
ranges between $0.5$~eV and $1.0$~eV.\cite{Bard1980} In our calculations
we have used intermediate value of $0.71\,\mbox{eV}$.\cite{Szroeder2011}

Dependence of the cathodic reaction rate on the electrode potential
$k_{c}(V)$ is given by the integral \cite{Memming2001,Szroeder2011}
\begin{equation}
k_{c}\propto\int\left[1-f(E,V)\right]\,\mbox{DOS}(E,V)\, P(E)\, dE.\label{Eq_rate_const}
\end{equation}
Here, $\mbox{DOS}(E,V)=\mbox{DOS}(E-eV)$, and $f(E,V)=f(E-eV)$,
where $f(E)$ is the Fermi--Dirac distribution. To determine the position
of electron bands of graphene electrode in relation to the Gaussian
distribution of energy levels of the reduced form, vacuum energy has
been chosen as a reference. The value of $E_{F}$ in relation to the
vacuum energy is equal to the work function, which has been determined
experimentally for mono- and bilayer graphene using Kelvin probe force
microscopy giving the $E_{F}${[}\textit{vs}.~vacuum{]} $=-4.6$~eV.\cite{Yu2009}

For the Fe(CN)$_{3}^{3-/4-}$ redox couple, we used the $E_{F\mathrm{,redox}}${[}\textit{vs}.~vacuum{]}
value determined from the half wave potential obtained by cyclic voltammograms.
Measurements carried out on epitaxial graphene and HOPG give value
ranging from $V_{1/2}=-0.025$~V~\textit{vs.}~Ag/AgCl to $V_{1/2}=0.268$~V~\textit{vs.}~Ag/AgCl.
\cite{Szroeder2014} According to Ref.~\onlinecite{Trasatti1974},
zero potential of the Ag/AgCl reference electrode is shifted in relation
to the vacuum potential by $-5.04\,\mbox{V}$. Assuming the half wave
potential value to be $0.20\,\mbox{V}$, we found the value of $E_{F\mathrm{,redox}}${[}\textit{vs}.~vacuum{]}
$=-4.84$~eV. Thus, we have assumed in our model that the Fermi level
of the Fe(CN)$_{3}^{3-/4-}$ redox couple is shifted in relation to
the Fermi level of the graphene by $-1.27$~eV.

\section{Results and Discussion}

\begin{figure*}
\includegraphics[width=1\textwidth]{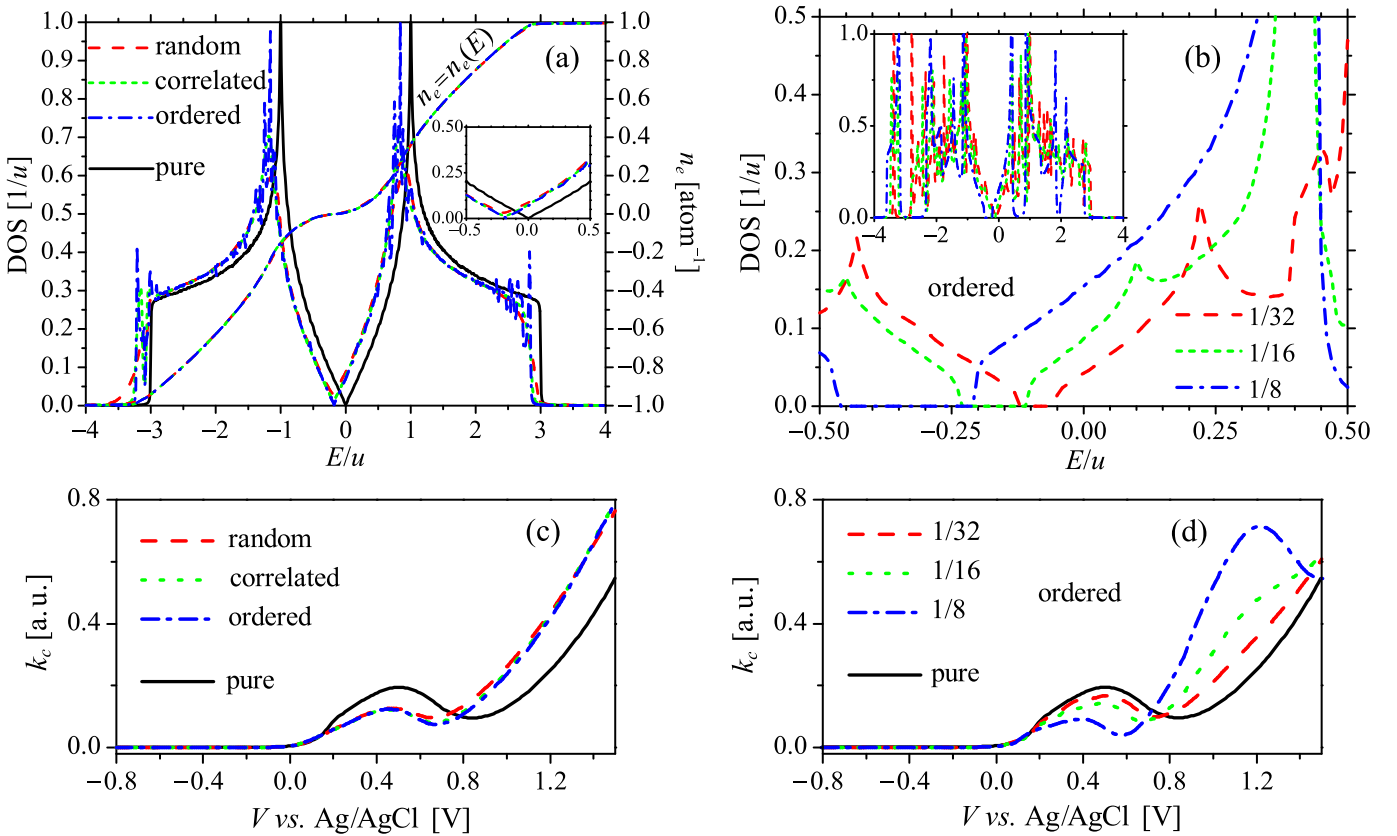}

\caption{(Color online) (a), (b) Density of states (DOS) for (a) potential
in Fig.~\ref{Fig_potential}(a) and for (b) short-range Gaussian
\textcolor{black}{potential $V=Ue^{-\mathbf{\mathrm{\mathit{r}}}^{2}/2\xi^{2}}$
(with potential height $U=-2u$ and effective potential radius $\xi=0.5a$}),
where (a) $3.125\%$ of random, correlated, or ordered adatoms occupy
hollow sites, while (b) $3.125\%$ (stoichiometry $1/32$), $6.25\%$
($1/16$), and $12.5\%$ ($1/8$) of ordered adatoms reside on top
sites. (c), (d) Rate constant ($k_{c}$) of the cathodic reaction
of oxidation of Fe(CN)$_{3}^{4-}$ at mono-layer graphene electrode
for respective (a), (b) DOS. Insets in (a) and (b) show the same as
in the main panels, but with another scales. As a reference, DOS (a)
and rate constant (c), (d) for pure graphene electrode are shown.}

\label{Fig_DOS+rate_constant} 
\end{figure*}

As it was mentioned in Secs.~I and II, we consider potassium as an
example of adsorbate in the present study. The most energy favorable
adsorption sites for K dopants in graphene are \textit{H}-type sites
as listed in Table~\ref{Table_potassium}. Therefore results obtained
at potential in Fig.~\ref{Fig_potential}(a) and \textit{H}-type
sites are appropriate for K adsorbed graphene first of all. Results
obtained for \textit{B}- and \textit{T}-type sites can be associated
with K adsorbate in a model assumption for revealing manifestation
of configurational effects in electron transport. 

Figure~\ref{Fig_DOS+rate_constant}(a) shows the DOS and the electron
density $n_{e}=n_{e}(E)$ for graphene with $n_{\mathrm{K}}=3.125\%$
of random, correlated, and ordered potassium adatoms, which are described
by the scattering potential in Fig.~\ref{Fig_potential}(a) and are
distributed over the \textit{H}-type adsorption sites. DOS-curves
for \textit{B}- and \textit{T}-type are similar to those shown in
Fig.~\ref{Fig_DOS+rate_constant}(a) with difference that Fermi level
in case of \textit{T} sites is shifted more far with respect to $E=0$
to the (left) side of negative \textit{$E<0$}---energies of holes
in our denotations. The Dirac (neutrality) point shifts towards negative
energies (gate voltage) due to electron (\textit{n}-type) doping dictated
by the asymmetry (negativity) of the scattering potential. The calculated
DOS-curves in Fig.~\ref{Fig_DOS+rate_constant}(a) for random, correlated,
and ordered K adatoms are similar with two differences take place
in the case of ordering: (i) peaks (fluctuations), which appear close
to $E/u\approx-3$ at correlation, begin to manifest themselves in
all energy interval (weakly away from the regions of the van Hove
singularities and $|E/u|\approx3$, but stronger close to them); (ii)
at a Fermi energy level, the DOS drops to zero (but even small band
gap does not open). Appearing of the peaks (fluctuations) in DOS is
due to the periodicity of the scattering-potential distribution describing
ordered positions of adatoms on the sites of interstitial {[}Figs.~\ref{Fig_Ordering}(left)
and \ref{Fig_Ordering}(center){]} or substitutional {[}Fig.~\ref{Fig_Ordering}(right){]}
graphene-based superstructure. Additional calculations\cite{Radchenko_NOVA2014}
{[}see also inset in Fig.~\ref{Fig_DOS+rate_constant}(b){]} show
that the peaks become stronger and even transform into discrete energy
levels with broadening as impurity concentration and/or periodic potential
increase. 

Positioning of ordered adatoms on the \textit{T}-type sites {[}Fig.~\ref{Fig_Ordering}(right){]}
makes possible band gap opening, which is clearly seen in Fig.~\ref{Fig_DOS+rate_constant}(b)
for a strongly short-range, \textit{e.g.}, Gaussian potential with
very small effective potential radius \textcolor{black}{$\xi<a$}.
The band gap is induced by the periodic potential leading to the ordered
distribution of adatoms directly above the C atoms belonging to the
same sublattice, thus breaking of symmetry of two graphene sublattices.
Note that adatoms on \textit{T} sites act as substitutional point
defects---impurities or vacancies---which can also induce the band
gap opening if they are distributed orderly \cite{Cheianov2010,Park2008,Martinazzo2010,Casolo2011,Radchenko_PLA2014}
or belong to the same sublattice even being randomly located. \cite{Lherbier2013,Pereira2008}
However, we did not observe the band gap appearing if ordered adatoms
reside on \textit{H} and \textit{B} adsorption sites (thereby act
as interstitial dopants) as it is reported by Cheianov \textit{et
al}. for adatoms occupying \textit{H} \cite{Cheianov2009SSC} and
\textit{B} \cite{Cheianov2009PRB} sites. We attribute the lack of
the band gap opening (when ordered adatoms occupy \textit{H} and \textit{B}
sites) to the absence of the breaking of global lattice symmetry in
these cases. 

Obtained densities of electronic states enter into Eq.~(\ref{Eq_rate_const})
and thereby enable us to calculate the electrode-potential-dependent
rate constants at various adatomic configurations as well as concentrations.
Figure \ref{Fig_DOS+rate_constant}(c) demonstrates the rate constant
of the reaction of oxidation of ferrocyanide ions, $k_{c}$, at graphene
electrode with potassium impurity as a function of electrode potential.
At a potential of about $-0.1$~V~\textit{vs.}~Ag/AgCl, increase
of the cathodic reaction rate is observed. Contrary to metallic electrodes,
the increase of the $k_{c}$ is not monotonic. In the range of calculated
potentials, the plot of $k_{c}$~\emph{vs.}~$V$ has a hump. The
local minimum appears within the electrochemical window of water,
\textit{i.e.} within the range $0.6\mathrm{-}0.8$~V~\textit{vs.}~Ag/AgCl.
The monotonic increase is observed at the positive electrode potentials
beyond the water window ($V>0.8$~V~\textit{vs.}~Ag/AgCl). Difference
between the $k_{c}$~\emph{vs.}~$V$ plots calculated for the pure
graphene and graphene with K adatoms is seen clearly. However, as
well as the shape of DOS-curves close to the Dirac point {[}Fig.~\ref{Fig_DOS+rate_constant}(a){]},
the shape of $k_{c}$-plots of impure graphene {[}Fig.~\ref{Fig_DOS+rate_constant}(c){]}
is only negligibly affected by the impurity configuration. Generally,
K-impurity slow down the reaction kinetics in the electrode potentials
in the range of water window. At the pure graphene electrode, the
maximum of the hump is located at potential of $0.503\,\mbox{V}$~\textit{vs.}~Ag/AgCl
($k_{c}=0.194$~a.u.), whereas at graphene with K adatoms in the
concentration of $3.125\%$ the maximum is observed at lower potential
of $0.453\,\mbox{V}$~\textit{vs.}~Ag/AgCl ($k_{c}=0.123$~a.u.).
The same applies to the position of the local minimum, which is downshifted
in graphene with K adatoms by $0.17\,\mbox{V}$.

Contrary to the impurity configuration, the concentration of adatoms
influences strongly the $k_{c}$ {[}Fig.~\ref{Fig_DOS+rate_constant}(d){]}
similarly to the influence on the DOS {[}Fig.~\ref{Fig_DOS+rate_constant}(b){]}.
With increasing adatomic concentration, the position of the local
maximum shift towards lower potentials from the value of $0.503$~V~\textit{vs.}~Ag/AgCl
($k_{c}=0.123$~a.u.) for pure graphene to $0.393$~V~\textit{vs.}~Ag/AgCl
($k_{c}=0.091$~a.u.). Also the local minimum shifts down form the
potential value of $0.843$~V~\textit{vs.}~Ag/AgCl to $0.393$~V~\textit{vs.}~Ag/AgCl
to $0.573$~V~\textit{vs.}~Ag/AgCl for pure graphene and graphene
with adatomic concentration of $12.5\%$, respectively. In a part
of the $k_{c}$ plot in the range of higher potentials ($V>0.8$~V~\textit{vs.}~Ag/AgCl)
an additional hump is apparent at $12.5\%$ of adatoms.

Assuming $0.2$~V~\textit{vs.}~Ag/AgCl as a standard electrode
potential (when the rates of both cathodic and anodic reactions are
equal), in Table~\ref{Table_stand_rate_const} we compare values
of the standard rate constant, $k_{s}$, for electron transfer between
the Fe(CN)$_{3}^{3-/4-}$ redox couple and impure graphene at weakly
long-range scattering potential in Fig.~\ref{Fig_potential}(a) and
strongly short-range Gaussian potential \textcolor{black}{$V=Ue^{-\mathbf{\mathrm{\mathit{r}}}^{2}/2\xi^{2}}$
with potential height $U=-2u$ and effective potential radius $\xi=0.5a$}.
While the adatomic configurations do not affect significantly the
shape of the $k_{c}$ plot, apparent differences in the $k_{s}$ at
different ranges of the scattering potential action are seen. When
the long-range potential is used, the electron transfer is moderately
suppressed by adatoms in random ($\approx$31\%), correlated ($\approx$33\%),
and ordered ($\approx$32\%) configurations as compared to the electron
transfer of pure graphene. On the other hand, weak dependence between
impurity concentration and the $k_{s}$ value is observed if the short-range
potential is used; quadrupling the adatomic content causes the decrease
of $k_{s}$ by only $\approx$23\%. Thus, the use of the long-range
potential more strongly suppresses the electron transfer kinetics.
It is worth noting that the best kinetics is observed at pure electrode.
Our findings are not compatible with experimental results obtained
in Ref.~\onlinecite{McCreery2008}. Discrepancies are probably
due to the hydrophobic properties of graphene.

\begin{table}
\caption{Standard rate constants, $k_{s}$, for electron transfer between graphene
electrode and ferro-/ferricyanide redox couple for different ranges
of the scattering-potential action, adatomic configurations and concentrations.}

\begin{tabular}{llll}
\hline 
Type of potential & Configuration  & Stoichiometry  & $k_{\mathrm{\mathit{s}}}$ {[}$\mbox{a.u.}${]}\tabularnewline
\hline 
\multirow{3}{*}{Long-range} & Random  & \multirow{3}{*}{1/32 (3.125\%) } & 0.0595\tabularnewline
 & Correlated  &  & 0.0571\tabularnewline
 & Ordered &  & 0.0586\tabularnewline
\hline 
\multirow{3}{*}{Short-range} & \multirow{3}{*}{Ordered } & 1/32 (3.125\%)  & 0.0791\tabularnewline
 &  & 1/16 (6.25\%)  & 0.0740\tabularnewline
 &  & 1/8 (12.5\%)  & 0.0610\tabularnewline
\hline 
 & Pure  & 0  & 0.0858\tabularnewline
\hline 
\end{tabular}

\label{Table_stand_rate_const} 
\end{table}

\begin{figure*}
\includegraphics[width=1\textwidth]{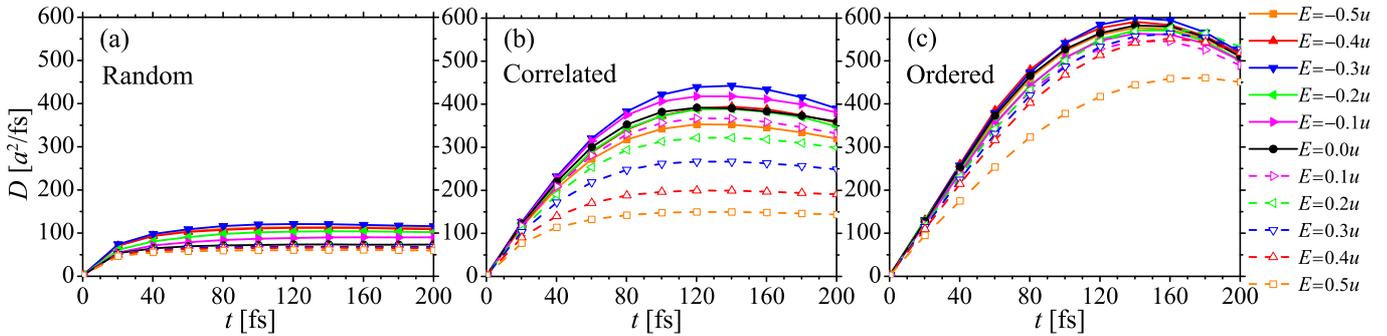}

\caption{(Color online) Time-dependent diffusivity within the energy range
$E\in[-0.5u,0.5u]$ for random (a), correlated (b), and ordered (c)
potassium adatoms located on hollow sites.}

\label{Fig_Diff} 
\end{figure*}

\begin{figure*}
\includegraphics[width=1\textwidth]{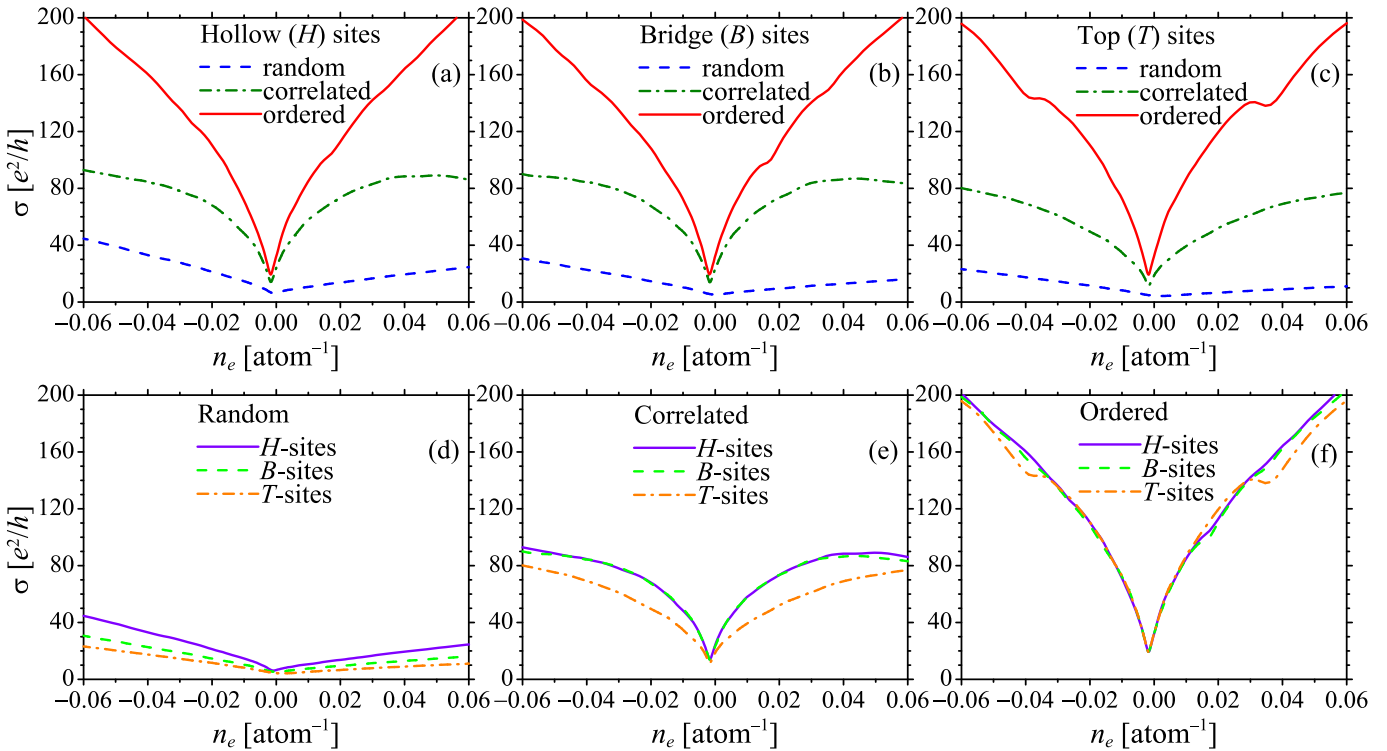}

\caption{(Color online) Conductivity \textit{vs}. the electron density for
$n_{\mathrm{K}}=3.125\%$ of random, correlated, and ordered potassium
adatoms occupying hollow (\textit{H}), bridge (\textit{B}), or top
(\textit{T}) adsorption sites. Curves in upper and lower figures are
the same, but grouped in a different way to distinguish configuration
effects induced by correlation or ordering from those caused by difference
in type of adsorption sites: \textit{H}, \textit{B}, or \textit{T}.}

\label{Fig_cond_hollow-bridge-top} 
\end{figure*}
\begin{figure*}
\includegraphics[width=1\textwidth]{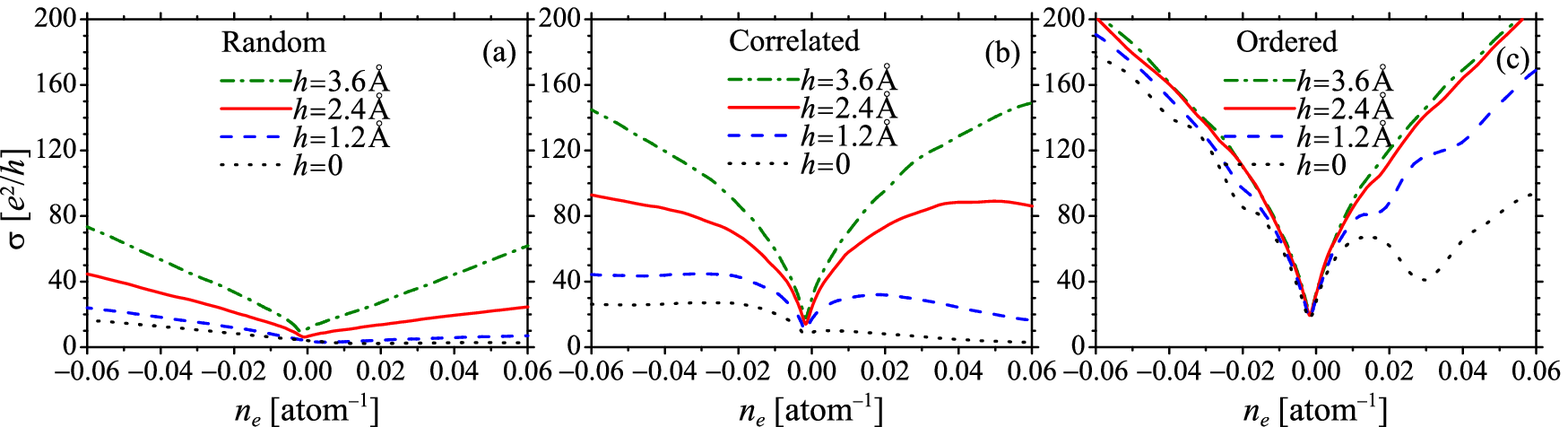}

\caption{(Color online) Electron-density-dependent conductivity for different
adsorption heights, \textit{h}, of $3.125\%$ of random (a), correlated
(b), and ordered (c) K adatoms resided on hollow sites.}

\label{Fig_cond_height} 
\end{figure*}

In contrast to the case of randomly-arranged adatoms, when steady
diffusive regime is reached for a relatively short time {[}Fig.~\ref{Fig_Diff}(a){]},
in case of their correlation and especially ordering, a quasi-ballistic
regime is observed during a long time as it is shown in Figs.~\ref{Fig_Diff}(b)
and (c). This (quasi-ballistic) behavior of diffusivity, $D(t)$,
indicates a very low scattered electronic transport, at which maximal
value of $D(t)$ is substantially higher for correlated and much more
for ordered adatoms as compared with their random distribution. If
the diffusive regime is not completely reached, the semiclassical
conductivity, $\sigma$, cannot be in principle defined. However,
we extracted $\sigma$ for the case of ordered adatoms using the highest
$D(t)$ when quasi-ballistic behavior turns to a quasi-diffusive regime
with an almost saturated diffusivity coefficient.

Figure~\ref{Fig_cond_hollow-bridge-top} represents calculated conductivity
($\sigma$) as a function of electron ($n_{e}>0$) or hole ($n_{e}<0$)
concentration, $\sigma=\sigma(n_{e})$, for different positions (\textit{viz.
H}, \textit{B}, and \textit{T}) and distributions (\textit{viz.} random,
correlated, and ordered) of adatoms in graphene. For visual convenience,
we arranged the same (nine) curves in two groups: Figs.~\ref{Fig_cond_hollow-bridge-top}(a)--(c)
demonstrate how correlation and ordering affect the conductivity for
each of \textit{H}, \textit{B}, and \textit{T} adsorption types, while
Figs.~\ref{Fig_cond_hollow-bridge-top}(d)--(f) exhibit how these
three types of sites influence on the conductivity for each of random,
correlated (with maximal correlation lengths as listed in Table \ref{Table_correlation}),
and ordered adatomic distributions. The conductivity exhibits linear
or nonlinear (\textit{viz.} sublinear) electron-density dependencies.
The linearity of $\sigma=\sigma(n_{e})$ takes place at randomly-distributed
potassium adatoms and indicates dominance of the long-range contribution
to the scattering potential, while sublinearity occurs at nonrandom
(\textit{viz.} correlated and ordered) positions of K adatoms and
is indicative of the dominance of short-range component of the scattering
potential. This is in accordance with many previous studies (see,
\textit{e.g.}, Ref.~\cite{Radchenko_PRB2012} and references therein)
in which pronounced linearity and sublinearity of $\sigma=\sigma(n_{e})$
are observed for long-range scattering potential (appropriate for
screened charged impurities ionically bond to graphene) and short-range
potential (appropriate for neutral covalently bond adatoms), respectively.
These results illustrate manifestation of contrasting scattering mechanisms
for different spatial distributions of metallic adatoms.

One can see from Figs.~\ref{Fig_cond_hollow-bridge-top}(a)--(c)
that conductivities for correlated ($\sigma_{\mathrm{cor}}$) and
ordered ($\sigma_{\mathrm{ord}}$) adatoms are dozens of times enhanced
as compared with case of randomly-distributed adatoms ($\sigma_{\mathrm{rnd}}$).
These enhancements ($\sigma_{\mathrm{cor}}/\sigma_{\mathrm{rnd}}$
and $\sigma_{\mathrm{ord}}/\sigma_{\mathrm{rnd}}$) depends on electron
density and type of adsorption sites. It is easy to determine from
Figs.~\ref{Fig_cond_hollow-bridge-top}(a)--(c) that the ratio $\sigma_{\mathrm{cor}}/\sigma_{\mathrm{rnd}}$
ranges as 2\,$\lesssim$\,$\sigma_{\mathrm{cor}}^{H}(n_{e})/\sigma_{\mathrm{rnd}}^{H}(n_{e})$\,$\lesssim$\,5,
2\,$\lesssim$\,$\sigma_{\mathrm{cor}}^{B}(n_{e})/\sigma_{\mathrm{rnd}}^{B}(n_{e})$\,$\lesssim$\,6,
and 3\,$\lesssim$\,$\sigma_{\mathrm{cor}}^{T}(n_{e})/\sigma_{\mathrm{rnd}}^{T}(n_{e})$\,$\lesssim$\,7;
while $\sigma_{\mathrm{ord}}/\sigma_{\mathrm{rnd}}$ ranges as 3\,$\lesssim$\,$\sigma_{\mathrm{ord}}^{H}(n_{e})/\sigma_{\mathrm{rnd}}^{H}(n_{e})$\,$\lesssim$\,8,
3\,$\lesssim$\,$\sigma_{\mathrm{ord}}^{B}(n_{e})/\sigma_{\mathrm{rnd}}^{B}(n_{e})$\,$\lesssim$\,9,
and 4\,$\lesssim$\,$\sigma_{\mathrm{ord}}^{T}(n_{e})/\sigma_{\mathrm{rnd}}^{T}(n_{e})$\,$\lesssim$\,15
(here, superscripts denote types of adsorption sites). As follows
from Figs.~\ref{Fig_cond_hollow-bridge-top}(d)--(f), in a random
adatomic state $\sigma_{\mathrm{rnd}}^{T}$\,$<$\,$\sigma_{\mathrm{rnd}}^{B}$\,$<$\,$\sigma_{\mathrm{rnd}}^{H}$,
while for correlated and ordered states, $\sigma_{\mathrm{cor}}^{T}$\,$<$\,$\sigma_{\mathrm{cor}}^{B}$\,$\approx$\,$\sigma_{\mathrm{cor}}^{H}$
and $\sigma_{\mathrm{ord}}^{T}$\,$\approx$\,$\sigma_{\mathrm{ord}}^{B}$\,$\approx$\,$\sigma_{\mathrm{ord}}^{H}$,
respectively. That is why the highest increase of $\sigma$ due to
correlation ($\sigma_{\mathrm{cor}}/\sigma_{\mathrm{rnd}}$) or ordering
($\sigma_{\mathrm{ord}}/\sigma_{\mathrm{rnd}}$) takes place for the
\textit{T-}site bonding, followed by the \textit{B} sites, and then
the \textit{H} sites. The increasing of $\sigma$ due to adatomic
correlation or ordering is expected in a varying degree for any constant-sign
($V>0$ or $V<0$) scattering potential, but this is not the case
when the potential is sign-changing ($V\gtrless0$).\cite{Radchenko_PRB2012}

If adatoms are randomly-positioned on the (\textit{H}, \textit{B},
or \textit{T}) adsorption sites, the conductivity is dependent on
their type: $\sigma_{\mathrm{rnd}}^{H}$\,$>$\,$\sigma_{\mathrm{rnd}}^{B}$\,$>$\,$\sigma_{\mathrm{rnd}}^{T}$,
particularly $\sigma_{\mathrm{rnd}}^{H}$\,$\thickapprox$\,$2\sigma_{\mathrm{rnd}}^{T}$
{[}Fig.~\ref{Fig_cond_hollow-bridge-top}(d){]}. Here, the differences
in $\sigma$ are caused by different values of on-site potentials
for these three types of adatomic positions although the same potential
profile $V=V(r)$ {[}Fig.~\ref{Fig_potential}(a){]} is used for
them. The stronger (weaker) on-site potential $V_{i}$ corresponds
to the smaller (larger) distance $r$ from the given graphene-lattice
site $i$ to the nearest adsorption site, which is more close (distant)
for the \textit{H} (\textit{T}) type, followed by the \textit{B} type,
and then the \textit{T} (\textit{H}) type. If adatoms are correlated,
the conductivity is dependent on whether they act as interstitial
(\textit{H} or \textit{B} sites) or substitutional (\textit{T} sites)
atoms: $\sigma_{\mathrm{cor}}^{H}$\,$\approx$\,$\sigma_{\mathrm{cor}}^{B}$\,$>$\,$\sigma_{\mathrm{cor}}^{T}$
{[}Fig.~\ref{Fig_cond_hollow-bridge-top}(e){]}, which can be attributed
to the values of maximal correlation lengths (Table~\ref{Table_correlation})
defining correlation degree for \textit{H}, \textit{B} and \textit{T}
sites, $r_{0_{\max}}^{H}$\,$=$\,$r_{0_{\max}}^{B}$\,$>$\,$r_{0_{\max}}^{T}$.
Finally, if adatoms form ordered superstructures (superlattices) with
equal periods (Fig.~\ref{Fig_Ordering}), the conductivity is practically
independent on the adsorption type (especially for not very high charge
carrier densities): $\sigma_{\mathrm{ord}}^{H}$\,$\approx$\,$\sigma_{\mathrm{ord}}^{B}$\,$\approx$\,$\sigma_{\mathrm{ord}}^{T}$
{[}Fig.~\ref{Fig_cond_hollow-bridge-top}(f){]}. 

In our model, increase (or decrease) of adatomic elevation over the
graphene surface results to more weak (or strong) scattering-potential
amplitude, \textit{i.e.} physically it means more weak (or strong)
regime of electron scattering on charged impurity adatoms. Although
the values of adsorption height, $h$, reported in the literature
for potassium do not disagree as much as for the adsorption energy
(see Table~\ref{Table_potassium}), for the model and calculation
completeness, we range $h$ in a wide interval (up to $h=3.6$~Å)
including an exotic case of $h=0$, when impurity atoms act as strictly
interstitial ones. Calculated curves representing the charge-carrier-density-dependent
conductivity for (random, correlated, and ordered) adatoms resided
on (the most favorable for potassium) hollow sites and elevated on
different $h$ are shown in Fig.~\ref{Fig_cond_height}. (Here, we
do not consider the cases of less favorable for potassium bridge and
top sites since it leads to qualitatively the same results.) As follows
from Figs.~\ref{Fig_cond_height}(a) and (b), at least for hole densities
($-n_{e}>0$), two (three) time increased or decreased $h$ for randomly-
or correlatively-distributed K-adatoms results to approximately two
(three) time enhanced or reduced $\sigma$, respectively. Thus the
conductivity approximately linearly scales with adsorption height
of random or correlated adatoms, $\sigma(h)\propto h$. However, for
ordered potassium adatoms, the $\sigma$ remains practically unchanged
with varying of $h$ in the realistic range of adsorption heights
(see Table~\ref{Table_potassium}) and even in all range at issue
($0\leqslant h\leqslant3.6$ Å) for hole densities {[}Fig.~\ref{Fig_cond_height}(c){]}.
We attribute this to the dominance of short-range scatterers in case
of their ordered state as it was mentioned above. Indeed, the Gaussian
fitting\textcolor{black}{{} for} the scattering potential in Fig.~\ref{Fig_potential}(a)
yields the effective potential radius \textcolor{black}{$\xi=2.21a$},
\textcolor{black}{which is commensurable with quantities of adsorption
heights $h$ at issue (and even less than }$h=3.6$~Å). 

In conclusion of this section, note that our numerical calculations
of conductivity in Figs.~\ref{Fig_cond_hollow-bridge-top} and \ref{Fig_cond_height}
agree with experimentally observed features of $\sigma=\sigma(n_{e})$
in potassium-doped graphene:\cite{Chen2008,YanFuhrer2011} (i) asymmetry
in the conductivity for electrons versus holes (which, however, can
be weakened and even totally suppressed due to the spatial correlation
or especially ordering of adatoms as well as increasing of their adsorption
height), (ii) shifting of minimum conductivity at a charge neutrality
point to more negative gate voltage, (iii) linearity or sublinearity
of conductivity at lower or higher gate voltage, respectively, and
(iv) increase in conductivity due to correlation in the positions
of adatoms that was also sustained theoretically\cite{Li2011} within
the standard semiclassical Boltzmann approach in the Born approximation.
A significant sublinear behavior of electron-density-dependent conductivity
and its saturation for very high densities at the spatial correlations
among the charged impurity locations in contrast to the strictly linear-in-density
graphene conductivity for uncorrelated random charged impurity scattering
{[}Figs.~\ref{Fig_cond_hollow-bridge-top}(a)--(e) and \ref{Fig_cond_height}(a)--(b){]}
is also in agreement with theoretical findings in Refs.~\onlinecite{Li2011}
and \onlinecite{Li2012}.

\section{Conclusions}

By employing numerical calculations, we systematically studied the
effects of different (random, correlated, and ordered) spatial configurations
of potassium adatoms onto high-symmetry {[}hollow- (\textit{H}), bridge-
(\textit{T}), and top-type (\textit{T}){]} adsorption sites with various
elevations over the graphene sheet on its electron transport and electrochemical
properties to ascertain correlation between them. We conclude as follows. 

(i) The charge carrier density dependence of the conductivity is indicative
of dominance of long-range scattering centers for their random spatial
distribution, while short-range scatterers dominate for their correlated
and ordered states. This demonstrates manifestation of contrasting
scattering mechanisms for different spatial distributions of metallic
adatoms. 

(ii) A band gap may be opened only if ordered adatoms act as substitutional
atoms (\textit{i.e.} reside on \textit{T}-type sites) due to the breaking
of graphene lattice point symmetry, while there is no band gap opening
for adatoms acting as interstitial atoms (\textit{i.e.} occupying
\textit{H}- or \textit{B}-type sites). 

(iii) If adatoms are randomly-positioned on the \textit{H}, \textit{B}-
or \textit{T} sites, the conductivity is dependent on their type:
$\sigma_{\mathrm{rnd}}^{H}$\,$>$\,$\sigma_{\mathrm{rnd}}^{B}$\,$>$\,$\sigma_{\mathrm{rnd}}^{T}$.
For spatially-correlated adatoms, the conductivity is dependent on
whether they act as interstitial or substitutional atoms: $\sigma_{\mathrm{cor}}^{H}$\,$\approx$\,$\sigma_{\mathrm{cor}}^{B}$\,$>$\,$\sigma_{\mathrm{cor}}^{T}$.
If adatoms form ordered superstructures (superlattices) with equal
periods, the conductivity is practically independent on the adsorption
type (especially for low electron densities): $\sigma_{\mathrm{ord}}^{H}$\,$\approx$\,$\sigma_{\mathrm{ord}}^{B}$\,$\approx$\,$\sigma_{\mathrm{ord}}^{T}$.

(iv) Depending on electron density and type of adsorption sites, the
conductivity for correlated and ordered K adatoms is found to be enhanced
in dozens of times as compared to the cases of their random positions.
The correlation and ordering effects manifest stronger for adatoms
acting as substitutional atoms and weaker for those acting as interstitial
atoms. 

(v) The electron--hole asymmetry in the conductivity for randomly-positioned
adatoms weakens and even may be totally suppressed for correlated
and especially ordered ones as well as for increased of their adsorption
height. 

(vi) The conductivity dependence with adsorption height of random
or correlated adatoms scales approximately as $\sigma(h)\propto h$.
However, for ordered adatoms, $\sigma$ remains practically unchanged
with varying of $h$ within its realistic range.

(vii) Only slight suppress of electron transfer kinetics in electrolyte
at K-doped graphene electrode is revealed. Strong correlation between
the band gap in graphene and the shape of the electrode-potential
dependence of electrochemical rate constant is seen, when the strongly
short-range scattering potential during the electron transport in
graphene is used. At the same time, the influence of this potential
on the suppress of the standard electrochemical rate constant is much
weaker as compared to the case of the long-range electron-scattering
potential in graphene. Comparison of the electron transfer calculations
to experiment shows that the hydrophobicity of graphene is a key factor,
which suppresses the kinetics of heterogeneous electron transfer in
electrolyte at graphene electrode. 
\begin{acknowledgments}
Authors acknowledge the Polish--Ukrainian joint research project under
the agreement on scientific cooperation between the Polish Academy
of Sciences and the National Academy of Sciences of Ukraine for 2015--2017
(No.~793). The work was also partly supported by the project ``Enhancing
Educational Potential of Nicolaus Copernicus University in the Disciplines
of Mathematical and Natural Sciences'' as part of Sub-measure 4.1.1
Human Capital Operational Programme (Project No.~POKL.04.01.01-00-081/10).
T.M.R. thanks Igor Zozoulenko and Artsem Shylau for their shared experience.\end{acknowledgments}


\begin{thebibliography}{100}
\bibitem{Fitting} Here, we use the standard way of finding the best
fit: so-called chi-square minimization procedure within the Levenberg--Marquardt
algorithm combining the Gauss--Newton method and the steepest descent
method.
\end{thebibliography}
\end{document}